\documentclass[sigconf,10pt]{acmart}
\usepackage{comment}
\usepackage{algorithm}
\usepackage[noend]{algpseudocode}
\usepackage{amsmath}
\usepackage{amsfonts}
\usepackage{subfig}
\usepackage{multirow}
\usepackage{graphicx}
\usepackage{multirow}
\usepackage{enumitem}
\usepackage{nccmath}
\usepackage{xspace}
\usepackage[T1]{fontenc}
\usepackage{circledsteps}

\usepackage[normalem]{ulem}

\usepackage{listings}
\usepackage{color}

\definecolor{dkgreen}{rgb}{0,0.6,0}
\definecolor{gray}{rgb}{0.5,0.5,0.5}
\definecolor{mauve}{rgb}{0.58,0,0.82}

\graphicspath{{figures/}}

\newcommand{\citadel}{Citadel\xspace}

\newcommand{\zhangrev}[1]{\textcolor{black}{{#1}}}

\newcommand{\wangc}[1]{}
\newcommand{\yanc}[1]{}
\newcommand{\zhangc}[1]{}

\usepackage{titlesec}
\titleformat{\paragraph}[runin]{\bfseries}{\thesubsubsection.}{.4em}{}
\titlespacing*{\paragraph}{0pt}{.4em plus .2em minus .2em}{.6em plus .2em minus .2em}

\begin{document}

\copyrightyear{2021}
\acmYear{2021}
\setcopyright{acmcopyright}\acmConference[SoCC '21]{ACM Symposium on Cloud Computing}{November 1--4, 2021}{Seattle, WA, USA}
\acmBooktitle{ACM Symposium on Cloud Computing (SoCC '21), November 1--4, 2021, Seattle, WA, USA}
\acmPrice{15.00}
\acmDOI{10.1145/3472883.3486998}
\acmISBN{978-1-4503-8638-8/21/11}

\date{}

\title{\citadel: Protecting Data Privacy and Model Confidentiality for
Collaborative Learning}

\author{
Chengliang Zhang*, Junzhe Xia*, Baichen Yang*, Huancheng Puyang*, Wei Wang*,\\
Ruichuan Chen$^\dagger$, Istemi Ekin Akkus$^\dagger$, Paarijaat Aditya$^\dagger$, Feng Yan$^\ddagger$\\
} 
\affiliation{*Hong Kong University of Science and Technology \quad $^\dagger$Nokia Bell Labs \quad $^\ddagger$University of Nevada, Reno}





\begin{abstract}

Many
organizations own data but have limited machine learning expertise (\emph{data owners}).
On the other hand, organizations that have expertise need data
from diverse sources to train truly generalizable models (\emph{model owners}).
With the advancement of machine learning (ML) and its growing awareness,
the data owners would like to pool their data and collaborate 
with model owners,
such that both entities can benefit from the obtained models.
In such a collaboration, the data owners want to protect the privacy of its
training data, while the model owners desire the confidentiality of
the model and the training method that may contain intellectual properties.
Existing private ML solutions, such as federated learning and split
learning, cannot \emph{simultaneously} meet the privacy requirements of both data and model owners.

We present \citadel, a scalable collaborative ML system that protects both data and model privacy in untrusted infrastructures equipped with
Intel SGX. \citadel performs distributed training across multiple
\emph{training enclaves} running on behalf of data owners and an
\emph{aggregator enclave} on behalf of the model owner. \citadel
establishes a strong information barrier between these enclaves by
\emph{zero-sum masking} and \emph{hierarchical aggregation} to prevent
data/model leakage during collaborative training. Compared with existing
SGX-protected systems, \citadel achieves better scalability and
stronger privacy guarantees for collaborative ML. Cloud deployment with
various ML models shows that \citadel scales to a large number of enclaves
with less than 1.73X slowdown.




\end{abstract}

\settopmatter{printfolios=false}
\settopmatter{printacmref=false}

\maketitle
\renewcommand{\shortauthors}{Zhang et al.}



\section{Introduction}
\label{sec:intro}

Building high-quality machine learning (ML) services requires not only extensive ML
expertise in feature selection, model design, hyperparameter tuning and
testing, but also a large amount of high-quality training data from diverse sources. However, these two requirements are often challenging to be met simultaneously, for instance, in healthcare and financial industries. 
The siloed data is often held by multiple entities (e.g., hospitals and banks), which are known as \emph{data owners}.
Each data owner alone may not have sufficient data nor ML expertise to train high-quality ML models. 
Thus data owners would like to collaborate with each other as well as the 
ML solution provider (e.g., technology firms) to build an intelligent service. 
The solution provider, known as a \emph{model owner}, can provide data owners
with an API to access the trained model and charge per use, similar to the
prevalent ML-as-a-Service practices adopted in Amazon and
Google~\cite{sagemaker,predictionapi}. Examples include
hospitals collaborating with an IT firm to train a diagnostic
imaging model over their patients' data~\cite{WatsonHealth}, and banks pooling
data to train an advanced fraud detection model developed by a FinTech company~\cite{altexsoft}. 

A key requirement for such collaborative ML is to protect both \emph{data
privacy} and \emph{model confidentiality}. For data owners, protecting the data from
being revealed to external entities is critical for protecting its business
interests, and more often than not, a regulatory
requirement~\cite{CCPA,GDPR,CyberSecurityLaw}. For a model owner, the model is a
valuable intellectual
property~\cite{tramer2016stealing,orekondy2019knockoff,jia2018preserving}.
Revealing proprietary model details (e.g., architecture and weights) can potentially result in losing technological advances to its market competitors. 
To make matters worse, the exposure of the model
raises new security issues in training and inference phases, such as
backdoor attacks, membership inference, and model
inversion~\cite{hitaj2017deep,mandal2019privfl,bagdasaryan2020backdoor,fredrikson2015model}.

Prevalent solutions for collaborative ML, such as federated
learning~\cite{yang2019federated,mcmahan2016federated,kairouz2019advances} and
split learning~\cite{gupta2018distributed,vepakomma2018split}, perform
training without exposing the participants' data, thus protecting data
privacy. However, they fail to protect model confidentiality as the training
model needs to be shared fully or partially among participants (see
\S\ref{sec:private_ML}). 

Trusted hardware, such as Intel Software Guard Extensions (SGX)~\cite{sgx}, has been used to facilitate collaborative ML with privacy guarantees for both data and model owners. A common approach is to perform ML training inside a single SGX enclave, where all training data and the model are loaded~\cite{hynes2018efficient}. However, this approach does not scale to large models nor large training datasets, due to the restricted size of the enclave page cache (EPC) (a few hundreds of megabytes~\cite{weichbrodt2018sgx}) and the excessive cryptographic overheads associated with evicting EPC pages to the main memory (see \S\ref{sec:singe_enclave}). Other approaches~\cite{pingan,hunt2018chiron,quoc2020securetf} distribute training amongst multiple enclaves but operate under a weaker threat model, thus insufficient for protecting both data privacy and model confidentiality in collaborative ML (see \S\ref{sec:multi-enclave}).

In this paper, we present \citadel, a scalable ML system that
enables collaborative learning in untrusted infrastructures by distributing the training across multiple SGX enclaves, with strong privacy
guarantees for both data and model owners.
\citadel proposes a privacy preserving mechanism to divide SGX-based ML training into training and aggregating parts, making it possible to spin up a distributed cluster to accommodate voluminous multi-sourced data.

To provide these properties, we need to address the following challenges.
First, providing privacy to data owners without compromising
model confidentiality is not straightforward.
SGX is a vehicle to execute trusted ML code, but data owners have to access and review all code running in the enclave to trust it in the first place.
Unfortunately, sharing the ML code is against the model owner's interest because it risks exposing proprietary ML techniques (see \S\ref{sec:threat_model}).
This paradox makes existing solutions either assume that model owners are not interested in stealing data and refrain from sharing code with data owners~\cite{hunt2018chiron}, or ensure that all data owners review and unanimously agree on the code running in the enclave~\cite{hynes2018efficient,quoc2020securetf}.
To our best knowledge, there is yet a practical SGX solution that can cater to both data and model owners' privacy requirements simultaneously.

Second, scaling the system with both data privacy and model confidentiality constraints is difficult.
It is already unrealistic to train a state-of-the-art ML model with a large dataset on a single machine within a reasonable amount of time, and adopting SGX only makes the matter worse by introducing steep performance degradation.
Without scaling out, an SGX-based solution becomes impractical despite its security advantages.
Careful deliberation is required to effectively and efficiently partition workload across multiple enclaves, and to enable them to securely collaborate together.

Third, reaping the security benefits of SGX requires adaptation of ML workloads.
Modern software, including ML tool chains, is usually memory-hungry, often using techniques like multiprocessing and single instruction multiple data (SIMD) operations to speedup data processing.
However, such techniques cause harmful effects in the memory-restricted SGX context. 
Our benchmarks demonstrate that the overhead of creating new processes and the excessive memory usage can easily overtake the benefits of parallelization. 
As a result, besides being able to scale out, reducing SGX overheads within each enclave is critical.

We tackle these challenges with a novel system design in \citadel.
First, to earn data owners' trust while protecting a model owner's confidentiality, we introduce two approaches, \emph{zero-sum masking} and \emph{hierarchical aggregation}, to isolate code that is handling data and code that is handling the model, and run these two parts in separate enclaves.
Therefore, only the code that has direct access to data shall be shared with data owners to gain trust, while the model handling code remains private to a model owner.
Second, with data handling code singled out, we are able to set up multiple such enclaves concurrently to process data in parallel, and aggregate the intermediate results together to update the model.
We also delegate the attestation and secret distribution to a centralized service \textsc{Pal{\ae}mon}~\cite{gregor2020trust}, so that model and data owners do not have to manage trust and secrets across many enclaves.
Third, we employ techniques like hierarchical aggregation and  multi-threading with pre-compiled C libraries to make ML workloads adapt to SGX's memory constraints.
We have implemented \citadel atop SCONE~\cite{arnautov2016scone}, a secure Linux container framework facilitating confidential computing with SGX.
Our implementation supports common distributed ML approaches such as local-update SGD~\cite{lin2018don,wang2018adaptive,haddadpour2019local}, model
averaging~\cite{mcmahan2016federated,haddadpour2019local}, and relaxed synchronization
\cite{ho2013more,lian2015asynchronous,zhang2018stay}, so that existing ML training applications can migrate to \citadel with minimal efforts.
\citadel is open-sourced.
We evaluate \citadel with various ML workloads of different sizes on Azure, and confirm that it can effectively speed up training via more enclaves.
With 32 training enclaves, we are able to boost the throughput to 4.7X--19.6X compared with those running in a single enclave.
We also demonstrate that \citadel achieves the privacy guarantees without significant overhead by comparing with baselines, including Chiron~\cite{hunt2018chiron} and running \citadel natively without SGX.
\section{Collaborative ML and Threat Model}
\label{sec:col_ML}


In many application domains such as healthcare and finance, building a
high-quality ML model requires the participation of both model and data
owners. The model owner (e.g., a tech company or ML developer) has advanced ML
expertise but may not have access to diverse training datasets. On the other
hand, data owners (e.g., hospitals and retailers) have quality labeled
datasets, but any single one may not have enough data samples or ML
expertise to build a quality model. 
An ideal solution enables collaboration among data owners and the model owner,
such that the model developed by the
latter can be trained over the data owned by the former, while still
preserving data privacy and model confidentiality. 

\subsection{Entities in Collaborative ML}
\label{sec:collaborative_ML}


Collaborative ML typically involves three entities, a model owner, a number of
data owners, and a third-party infrastructure such as a public cloud for providing training resources. 
These entities have different goals in collaborative ML.

\paragraph{Data Owner.} Data is a critical asset containing personal or
business-sensitive information. In business-to-customer settings,
the protection of data privacy is further mandated by regulations
(e.g., GDPR~\cite{GDPR}). The violation of such
requirements can lead to hefty fines and serious legal
consequences~\cite{gdpr_fine}. Therefore, a data owner wants to protect its data from being exposed to other entities, including the model owner,
the cloud, and other data owners.

\paragraph{Model Owner.} For the model owner, protecting the confidentiality
of the training model (i.e., design and weights) is a top requirement. First,
the model is a valuable intellectual property as its development
demands tremendous research and engineering
effort~\cite{tramer2016stealing,orekondy2019knockoff}. Protecting it 
helps maintain the model owner's technical advances and supports its
business as data owners would otherwise train the model by themselves. Second,
maintaining the model confidentiality is also a security requirement. Prior
work shows that sharing the model with (untrusted) participants poses
new threats that are hard to defend against, such as membership inference, model
inversion, and backdoor
attack~\cite{hitaj2017deep,mandal2019privfl,bagdasaryan2020backdoor,fredrikson2015model}. In 
security-critical applications such as fraud detection and spam filtering,
exposing the model details creates a wider attack surface as
adversaries can forge attacks to evade the model's defense mechanisms by
offline trial and error~\cite{papernot2017practical}.

In addition, the model
owner wants to conceal the \emph{ML training method}, such as
optimizer selection~\cite{ruder2016overview}, gradient
manipulation~\cite{zhang2019gradient}, and learning rate
schedule~\cite{zeiler2012adadelta}. These are critical to the
training performance. Selecting, combining, and configuring them require
extensive ML expertise, which are part of the model owner's intellectual
property.

\paragraph{Third-Party infrastructure.} Training complex ML models over large
datasets requires a large amount of computational resources, which model
and data owners may not have. Therefore, a common practice is to
rent a large number of virtual instances in a third-party cloud (e.g., Azure
and AWS) and perform distributed training across those instances.

\subsection{Threat Model}
\label{sec:threat_model}

Training is performed on a third-party cloud trusted by \emph{neither
data owners nor the model owner}. The cloud instances, including privileged
software like OS and hypervisor, are untrusted. Attacks can be performed by
the cloud provider or anyone with access to the OS and hypervisor. On the other hand, 
data and model owners trust the implementation of the trusted computing base (e.g., Intel SGX) and its attestation service. We also assume that the participants trust standard ML
frameworks like TensorFlow~\cite{abadi2016tensorflow} and
PyTorch~\cite{paszke2019pytorch}. These frameworks are under public scrutiny in open-source communities, and their security is orthogonal to this work.

We assume that data owners and the model owner are \emph{honest but curious}. 
They faithfully follow the training process specified by our system, as they have no incentive to hinder the training.
While doing so, however, data owners may collude with each other to obtain the
model and methods to perform training on
their own. Data owners also want to pry on each other's data to improve their competitiveness in the same business sector. The model owner, on the other hand,
may want to access the training data for illicit use. 


The model owner might theoretically engineer the model to preserve information from input data.
Such a vulnerability could happen as it stems from the concept of model confidentiality and the information retaining nature of ML models.
There are potential routes to address this issue.
First, \citadel can potentially export the trained model to a secure enclave directly for deployment, thus eliminating the model owner's access to plaintext model after training.
Second, \citadel can get a third-party entity, which has no conflict of interest (e.g., a government agency or a neutral authority), involved to verify that the
model is not maliciously engineered~\cite{code_audit}.

\paragraph{What does \citadel not protect?}
We do not
address denial-of-service
attacks, side-channel attacks~\cite{van2018foreshadow} and rollback attacks~\cite{parno2011memoir}, as there have been complementary mechanisms to prevent them~\cite{oleksenko2018varys,baumann2015shielding,krahn2018pesos,kim2019shieldstore}.
Besides, we do not protect against membership inference
and data extraction attacks from model owners, similar to the state-of-the-art federated learning approaches and other SGX-based solutions~\cite{shokri2017membership}.  

\section{Prior Arts and Their Insufficiency}
\label{sec:prior_arts}

In this section, we describe why prior work is insufficient to protect data privacy and model confidentiality for collaborative learning under the threat model introduced in \S\ref{sec:threat_model}.
We start by introducing existing solutions designed for different collaborative learning scenarios and explaining why they cannot be applied here.
We then introduce the SGX-based solutions, which are the most related to our work, and discuss their problems to achieve scalable collaborative learning.

\subsection{Solutions for Collaborative Learning}
\label{sec:private_ML}


Existing solutions to collaborative ML focus on data privacy without model confidentiality.

\paragraph{Federated learning (FL)} is a computing paradigm in which
multiple clients collaboratively train a shared model by uploading their local
updates to a central server for aggregation without exposing private training
data~\cite{yang2019federated,kairouz2019advances}. FL employs various
security measures to protect data privacy for clients, including secure multi-party computation (SMPC)~\cite{yang2019federated,mohassel2017secureml,mohassel2018aby,zhang2020privcoll,bonawitz2017practical,so2020turbo,du2004privacy},
differential privacy (DP)~\cite{shokri2015privacy,mcmahan2017learning,mcmahan2018general,geyer2017differentially}, and homomorphic encryption (HE)~\cite{phong2018privacy,liu2019secure,liu2018secure,cheng2019secureboost,zhang2020batchcrypt}.
However, FL is not designed to protect
model confidentiality as the training model is shared among all clients, each
being both a model owner and a data owner.

\paragraph{Split Learning (SL)} offers an alternative approach to
collaborative ML for training deep neural networks~\cite{vepakomma2018split,gupta2018distributed}. In SL, a neural network
is split into two parts at a certain layer, called a \emph{cut layer}. The
model owner releases the neural network up to the cut layer to data owners, while
keeping the remaining layers private. Data owners train the
network up to the cut layer with their private data and send the updates to a
central server, based on which the model owner trains the remaining network.
While this scheme preserves data privacy, the model confidentiality cannot be
fully protected, as the neural network up to the cut layer is shared
among data owners. In addition, the parameters of neural network 
are only accessible to data owners, meaning the model owner 
cannot access the complete network of the trained model.

Given that these collaborative learning solutions assume the actual training is distributed among data owners, they have fundamentally different system architectures and cannot be easily adapted to a scenario under our threat model.


\subsection{Intel SGX}
\label{sec:tee}

ML training requires access to both data and model. To protect their
privacy, the training must be performed at a secure place trusted by
both data and model owners. Trusted hardware like Intel SGX (Software Guarded Extensions~\cite{sgx}) offers a viable solution to create a trusted execution environment (TEE)
even if the underlying platform is untrusted.

Intel SGX  is one of the
most widely available hardware-assisted TEE, along with other
implementations such as ARM TrustZone~\cite{trustzone} and AMD Secure Memory
Encryption (SME)~\cite{AMD_SME}. It sets aside a protected memory region,
called an \emph{enclave}, within an application's address space. Code
execution and memory access in the enclave are strongly isolated from external
programs. The processor ensures that only code running in an enclave can access the data
loaded into it. External programs, including the operating system and
hypervisor, can only invoke code inside an enclave at the statically-defined entry
points. SGX also supports remote \emph{attestation}, which
allows a remote user to verify that the initial code and data loaded into an
enclave match a given cryptographic hash, hence creating trust that the enclave will perform the expected computation.

However, the enclave's hardware-protected confidentiality and integrity come
with a steep price for performance. First, as the host platform is untrusted,
copying between CPU and enclave memory must be protected to prevent memory bus
snooping. SGX uses memory encryption engine (MEE) to transparently encrypt and
decrypt data exchanges through memory bus, incurring 2X-3X performance
overhead than native executions~\cite{hunt2018ryoan}. Second, the performance
of an enclave is usually bounded by the enclave page cache (EPC) size, a hardware-protected memory region used to host the enclave pages.
The EPC is usually small, e.g., only 168MB in the most expensive Azure
confidential computing instance~\cite{azure_instance}. Any memory usage beyond
the EPC will cause enclave pages to evict to the unprotected main memory.
To ensure the confidentiality and integrity of the evicted EPC pages, SGX uses
symmetric key cryptography, which compounds to a large overhead
as the number of evictions increases. Such overhead can be mitigated by
optimizing code to avoid paging as much as possible. Third, because system
calls still need to be facilitated outside the enclaves, there is a substantial
context switching overhead. State-of-the-art SGX systems often avoid system
calls like I/O and threading~\cite{arnautov2016scone}.

\zhangrev{
Software Fault Isolation (SFI)~\cite{wahbe1993efficient} is a software
instrumentation technique for sandboxing untrusted modules, preventing codes from accessing others' secrets.
It has been adopted lately in Occlum~\cite{shen2020occlum} and Chancel~\cite{ahmad2021chancel} to provide certain levels of secret isolation.
However, SFI alone cannot provide strong security up to our threat model.
For example, a malicious model owner can run codes to compress users' data and export the unidentifiable data through SFI’s defined exit.
Consequently, we believe verification for codes with direct data access is essential. 
}


\subsection{Private ML with a Single SGX Enclave}
\label{sec:singe_enclave}

\begin{figure}[!tp]
    \centering
    \includegraphics[width=0.55\columnwidth]{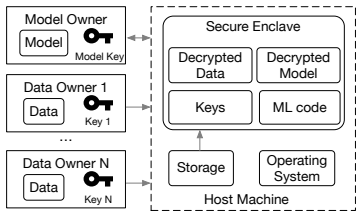}
    \vspace{-3mm}
    \caption{Illustration of a single-enclave that protects the
    confidentiality of both data and model.}
    \label{fig:enclave}
    \vspace{-6mm}
\end{figure}

One solution to private collaborative ML is to use a \emph{single} SGX
enclave attested by both data owners and the model owner (Fig.~\ref{fig:enclave}).\footnote{This design
is an extension to \cite{hynes2018efficient}, in which data owners also own
the training model, similar to the FL setting.} 
The training is
performed in a remote enclave running on an untrusted host (e.g., a cloud
server). Before the training begins, a data (or model) owner generates a private symmetric
key and uses it to encrypt the data (or model). The encrypted data and model are
then uploaded to an unprotected storage on the host, which does not have the key.
The host then creates an
enclave containing the \emph{agreed-upon ML code} by all secret (i.e., model and data) owners, and
lets them initiate attestation to ensure the integrity and correctness of the
initialized enclave. Once the attestation is successful, each secret owner uploads
its encryption key to the enclave over a TLS-protected channel, with which the
enclave can retrieve the encrypted data and model from the storage and decrypt
them. The
training starts once the data, model, and ML code are all loaded into the enclave. When the training completes, the model owner downloads the
model, and the enclave is destroyed along with the contained data.

\paragraph{Poor Scalability.}
Such a design, however, does not scale to a large training dataset. To illustrate
this problem, we characterize its performance with Azure's latest confidential
computing offering \texttt{DCsv2}~\cite{azure_instance}. We run experiments in
a \texttt{Standard\_DC8\_v2} instance, the largest in \texttt{DCsv2} with 8
vCPUs and 32~GB memory, of which 168MB is dedicated to an enclave's EPC. We
train AlexNet~\cite{krizhevsky2017imagenet} over images of size $32 \times 32
\times 3$ with TensorSCONE~\cite{kunkel2019tensorscone}, an SGX-optimized
version of TensorFlow v1.15. We then run the same training workload with the
unmodified TensorFlow outside the enclave. Note that, to speed up 
training on a single machine without accelerators, one common technique is to
configure a large batch size for increased parallelism and reduced iterations.
We evaluate the training epoch time (time needed to finish processing
the entire dataset) with varying batch sizes in SGX and the native environment,
respectively.

\begin{figure}
  \centering
  \label{fig:tf_single}
  \subfloat[Training epoch time and slowdown w.r.t. batch size]{
    \includegraphics[width=0.475\columnwidth]{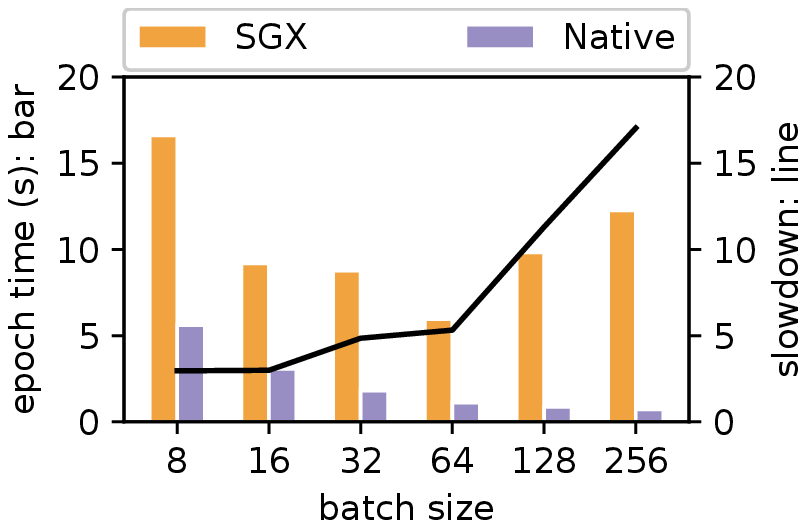}
    \label{fig:tf_single_slowdown}
  }\hfil
  \subfloat[Memory usage w.r.t. batch size]{
    \includegraphics[width=0.375\columnwidth]{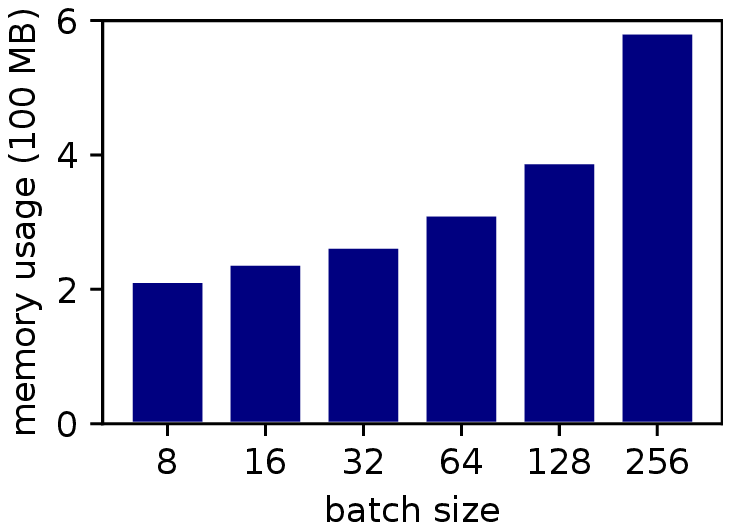}
    \label{fig:tf_single_mem}
  }
  \caption{Epoch time under SGX and native mode. The slowdown (line) represents the ratio between SGX and native time. Memory refers to active SGX memory.}
  \vspace{-.1in}
\end{figure}

When the
batch size is small (8 or 16), running in SGX is only 2.9X slower than running 
natively outside of the enclave, meeting the expected performance of
TensorSCONE~\cite{kunkel2019tensorscone} (Fig.~\ref{fig:tf_single_slowdown}). Such slowdown is mainly due to the
MEE encryption overhead but not EPC paging, as memory usage is barely over the EPC
size (Fig.~\ref{fig:tf_single_mem}). Further increasing the batch size leads to
more parallelism, which reduces the epoch time in the native
mode. This trend does not hold in SGX: as the batch size increases, the
epoch time first reduces but then surges rapidly, a consequence of frequent
EPC paging due to excessive memory usage beyond the EPC size
(Fig.~\ref{fig:tf_single_mem}). Therefore, one cannot expect to scale ML
training by configuring a large batch size in an enclave: with a batch size of 256, the slowdown with SGX is about 17X (Fig.~\ref{fig:tf_single_slowdown}).



\paragraph{Training Logic Exposure.}
Note that, in the single-enclave solution, the ML code must be shared and
agreed by all data owners to ensure that it contains no malicious code that
could harm their privacy (e.g., writing data to an external storage). However,
this sharing inevitably reveals the details of the model update logic (e.g., optimizer selection, learning rate scheduling, and gradient manipulation),
which the model owner may consider as intellectual property
(\S\ref{sec:col_ML}).

\subsection{Private ML with Multiple SGX Enclaves}
\label{sec:multi-enclave}

As model training in a single enclave does not scale, recent works propose distributed solutions with multiple enclaves. 
Notably, Ping An~\cite{pingan} augments
FL with SGX enclaves at the data owners' side for enhanced data privacy
while exploiting data parallelism, but the data owners can still
access the training model. Chiron~\cite{hunt2018chiron}, built atop
Ryoan~\cite{hunt2018ryoan}, ensures model confidentiality for ML-as-a-Service
providers with SGX enclaves, and supports running multiple training enclaves
in parallel. However, its design is based on the assumption that the model owner
(i.e., MLaaS provider) is not interested in harvesting data owners' data,
which may not be the case in collaborative ML.
\zhangrev{
The two approaches cannot be combined to complement each other, as Chiron places the training data in cloud servers, which is not allowed in Ping An.
}
SecureTF~\cite{quoc2020securetf} presents a modified TensorFlow to
support distributed training in multiple enclaves, but assumes that
model and data belong to the same entity, and hence cannot be applied to
collaborative ML.
PPFL~\cite{mo2021ppfl} adopts TEEs on mobile devices to train the final layers of a proprietary model in the context of cross-device federated transfer learning.
The data passes through a public base model, and the intermediate results are then fed into the TEEs holding secret model layers. In this design, only parts of the final model can remain private. Substantial information can also be inferred by a malicious model owner within the blackbox enclaves as they are not verified by data owners~\cite{phong2018privacy}.
Furthermore, mobile devices' constrained TEE memory greatly limits the ML model size and the training efficiency.
To our knowledge, a scalable collaborative ML system
that protects the privacy for both model and data owners is still lacking.

\section{\citadel Design}
\label{sec:solution}

We aspire to devise an ML system that not only preserves data and model privacy simultaneously, but also 
enables distributed training across multiple SGX enclaves.
To do so, we securely partition training workload, and make part of it replicable. 
\citadel achieves this by separating and isolating \emph{data handling code} and \emph{model handling code}.
The former can be shared with data owners to gain their trust, while the latter remains private to the model owner.
Afterwards, a \emph{barrier} has to be inserted between the two parts to ensure the model handling code cannot recover data owners' data with its private code.
With data handling codes isolated securely, \citadel is able to accelerate training through data parallelism.

\subsection{Design Overview}
\label{sec:design_overview}

\begin{figure}
    \centering
    \includegraphics[width=0.85\columnwidth]{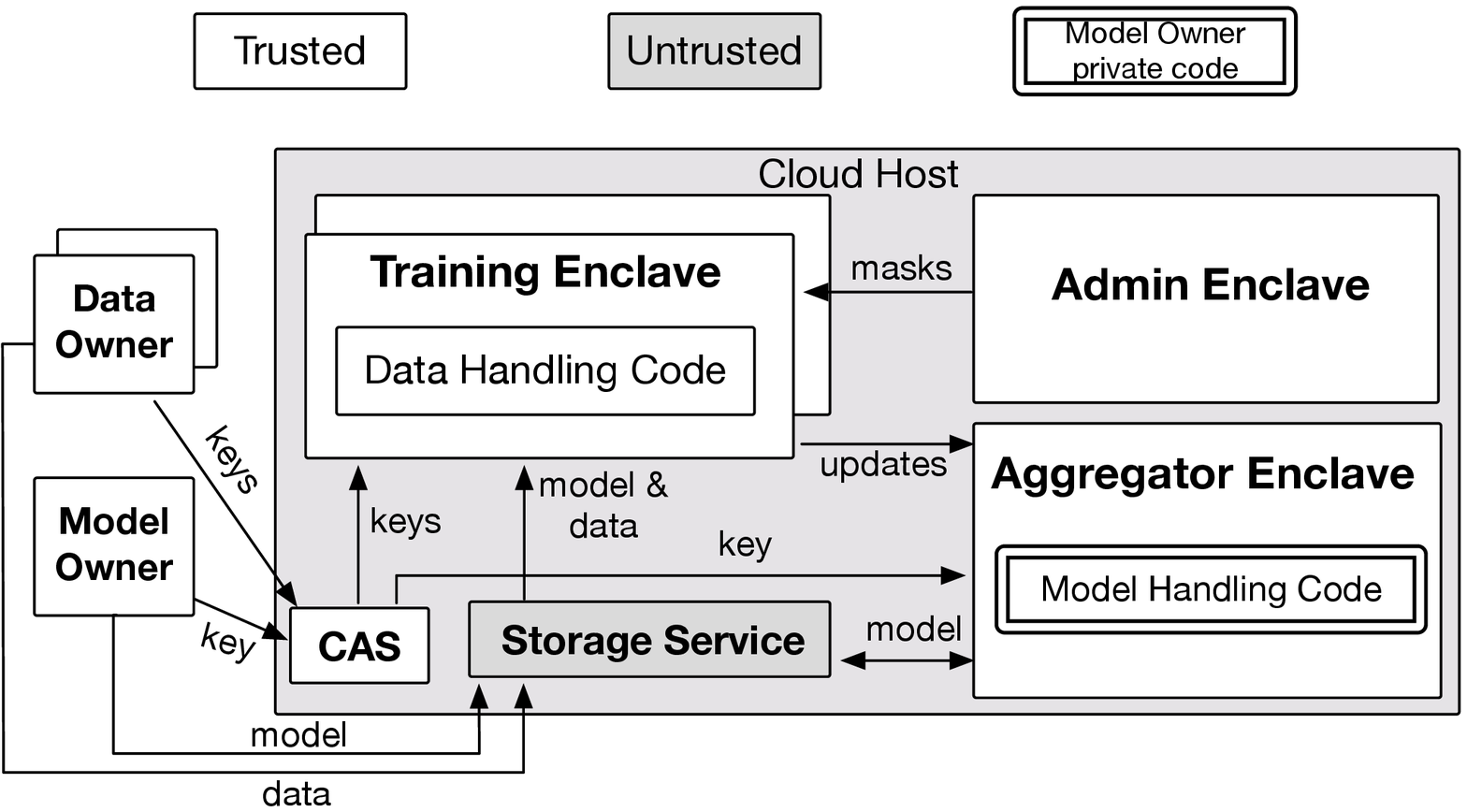}
    \caption{An overview of \citadel. All codes (except the
    model handling code) are open-sourced to gain data owner's trust, while the model handling codes remain private to model owner.
    }
    \label{fig:arch}
    \vspace{-.1in}
\end{figure}
Fig.~\ref{fig:arch} illustrates an architectural overview of our \citadel system. 
It facilitates collaborative ML in \emph{multiple} enclaves hosted in
untrusted infrastructure. These enclaves can run in a single or multiple cloud
instances. A data (or model) owner communicates with \citadel through a
\emph{client} running on a local machine. The client includes a \emph{verifier} which
the data (or model) owner uses to attest \citadel. It also provides a
\emph{key manager} with which the owner generates a symmetric encryption key and uses
it to encrypt data (or model). The client uploads the encrypted data (or model) to a
general cloud storage service. 
The storage itself does not need to be trusted since the secrets are encrypted. 
\citadel
launches multiple enclaves on behalf of data and model owners, establishes
trust between the enclaves and the owners via attestation (handled by
CAS, configuration and attestation service), and performs distributed training in those enclaves.  Specifically, \citadel runs
three types of enclaves:
\emph{training enclaves}, \emph{aggregator enclave}, and \emph{admin enclave}.

\paragraph{Training Enclave.}
In \citadel, each data owner has a dedicated training enclave, which
needs to be attested by
both the corresponding data owner(s) and the model owner to gain their trust. 
It takes private data as input,
runs data handling code (e.g., compute gradient updates) 
provided by the model owner, and generates model updates. 
As the code has direct access to the training data, it must be shared to and agreed by
the data owner. The code should reveal no model information that the model
owner wants to protect, such as hyper-parameter configuration and the
training model. Instead, it loads such information as \emph{environment variables}
and \emph{non-executable binary model files} from the storage service. 
Note that, it is not possible to inject
malicious code into the model files, as they are non-executable in the standard ML toolchains.
After the model updates are computed, the training enclave sends them to the aggregator enclave for global aggregation.
\citadel currently does not consider the scenario where models are too big for a single enclave. Such an extension could be achieved by either increasing EPC size with a specialized SGX card~\cite{chakrabarti2019scaling}, or applying existing model parallelism techniques to split large models~\cite{yugillis,huang2019gpipe,harlap2018pipedream}.

\paragraph{Aggregator Enclave.} 
\citadel launches an aggregator enclave on behalf of the model owner to run the model handling code. Each
training job has only one such enclave, and it is attested by the model owner only.
It collects aggregated
updates (\S~\ref{sec:isolating}) from all training enclaves and updates the model
for the next training iteration.
The updated model is encrypted and stored in storage service, so that the training enclaves
can start the next iteration after retrieving it. As the aggregator enclave
has no access to data from data owners,
the code running inside remains private to the model
owner. This prevents sensitive training methods developed by the model
owner from being revealed (e.g., learning rate schedule, optimizer selection,
gradient selection and manipulation), which are required for aggregation. 

\paragraph{Admin Enclave.}
\citadel launches an admin enclave for a training job, uses it to schedule 
the training workload and orchestrate the associated training and aggregator
enclaves. Code running inside an admin enclave (i.e., \emph{mask generator} described later in \S\ref{sec:isolating} and \emph{enclave scheduler})
are open-sourced for public access. The enclave itself is
attested by all model and data owners.
To facilitate communication between an enclave and external entities, \citadel
provides open-source utilities that run as part of \emph{admin code} in a training or
aggregator enclave. As the cloud host's network is
untrusted, communications inside \citadel are secured by TLS connections
with endpoints located inside the enclaves.

\paragraph{Attestation with CAS.}
In \citadel, a secret owner needs to attest multiple enclaves to ensure the
integrity and confidentiality of the data and code. The default SGX
attestation can be tedious as it attests one enclave at a time.
CAS (configuration and attestation service) offers a simplified solution to
secret management and attestation. CAS itself is open-sourced and runs in an
enclave, which the model and data owners can verify and attest. Once the
secret owners have established trust via CAS, they can delegate their
encryption keys to it, and instruct it on how to maintain their security including
which enclave can access which secrets and which code should run in
which enclave. CAS faithfully follows the specified security policy, attesting
each enclave on behalf of the data or model owners and supplying the enclave
with secrets once it is trusted. With the help of CAS, model and data owners
only have to initiate the attestation process once. \citadel employs 
\textsc{Pal{\ae}mon}~\cite{gregor2020trust}, a trust management service built
on top of SCONE~\cite{arnautov2016scone}, as its CAS system.

\paragraph{Fault Tolerance.}  
\citadel's training enclaves are \emph{stateless} by design, because model and data are all stored into and fetched from a storage system.
In case of training enclave failures, \citadel can easily  launch replacements and resume the training process via restarting the ongoing iteration.
The training progress is always checkpointed since the updated model is encrypted and stored into storage after each iteration.
If admin or aggregator enclaves fail, \citadel can also similarly restart the cluster and continue training.

\subsection{Separating Data and Model Handling}
\label{sec:isolating}

A key design in \citadel is to separate the model owner's ML code into
two parts: \emph{model handling code} and \emph{data handling code}. 
The model handling code computes the global model updates
based on the gradients received from the training enclaves.
As such, it concerns with potentially confidential methods and values.
\citadel runs the model handling code in the aggregator enclave.
In contrast, the data handling code is shared with data owners (i.e., open-sourced) to gain their trust.
It handles standard forward and backward propagation mechanisms, and has  access to data owners' private data.

This separation provides the model owners with model confidentiality: \citadel
ensures that the data owners only see \emph{placeholders} for the model and
hyperparameters, which are loaded dynamically into training enclaves after
attestation (\S\ref{sec:design_overview}). 
The secrets to load these values and replace the
placeholders are only shared after the attestation, such that model and
hyperparameters remain unknown to data owners.

On the other hand, this separation alone does not fully provide data privacy for data owners.
Although data owners can verify the open-source data handling code and ensure it does not leak private data,
prior work shows that a data owner's training data can still be inferred accurately from computed
gradients~\cite{phong2018privacy}. \citadel addresses this problem to protect data privacy with two methods:
First, data owners do not receive intermediate models from the model owner, such that they cannot pry into other data owners' data.
Second, a \emph{barrier} is inserted between the training enclaves and the aggregator
enclave, so that the model owner only receives \emph{aggregated} updates but not the raw updates from any individual training enclave.
Specifically, we propose two mechanisms for such a barrier: 
\emph{zero-sum masking} and \emph{hierarchical aggregation}.

\subsubsection{Zero-Sum Masking.}
\label{sec:masking}
Zero-sum masking, originally proposed for federated learning (FL)
as a way to implement secure aggregation~\cite{bonawitz2017practical}, allows
data owners to collectively generate masks and apply them to their individual
updates before sending them to the aggregator. The masks are generated in such a
way that they are canceled out when summed up, so that the 
updates are correctly aggregated. Since the aggregator does not have access to
individual masks, it cannot learn the raw gradients from any data owner.

Inspired by such a concept, we propose a new zero-sum masking scheme for
\citadel. Compared with the FL setting, where the masks have to be generated
among distributed data owners, TEE enables us to execute code that is
verified and trusted by the involved parties, so we can opt to a \emph{centralized}
mask generation approach.

\renewcommand{\lstlistingname}{Pseudocode}

\begin{lstlisting}[language=Python, caption=\zhangrev{Training Enclave.}, float=tp, floatplacement=tbp]
# download, decrypt and, process data with data owner's key
train_data = download_decrypt_data(data, os.environ[DATA_KEY])
train_x, train_y = preprocess_data(train_data)
# download, decrypt, and load model with model owner's key
model = download_decrypt_model(os.environ[MODEL_KEY])
TH, ml_toolchain.load_model(model)
# training
gradients = ml_toolchain.train(train_x, train_y, os.environ[BATCH_SIZE])
# request and apply mask, send only masked gradients
mask = citadel.get_mask()
masked_gradients = mask + gradients
citadel.send(masked_gradients)
\end{lstlisting}

\begin{lstlisting}[language=Python, caption=Admin Enclave., float=tp, floatplacement=tbp]
# initiate sum and masks
sum = 0, masks = []
# generate N - 1 random masks
for i in range(N - 1):
  mask_i = rand.generate_mask(shape)
  masks.append(mask_i)
  sum += mask_i
# add the last mask so that the total is 0
masks.append(-sum)
# distribute masks to training enclaves
citadel.distribute_masks(masks)
\end{lstlisting}

\begin{lstlisting}[language=Python, caption=\zhangrev{Aggregator Enclave.}, float=tp, floatplacement=tbp]
# collect and aggregate gradients from training enclaves 
all_masked_gradients = citadel.wait()
aggregated_gradients = numpy.sum(all_masked_gradients)
# download, decrypt, and load model with model owner's key
model = download_decrypt_model(os.environ[MODEL_KEY])
ml_toolchain.load_model(model)
optimizer = ml_toolchain.optimizer(os.environ[LEARNING_RATE_SCHEDULE])
# update model
clipped_gradients = ml_toolchain.clip(aggregated_gradients)
updated_model = ml_toolchain.apply(clipped_gradients)
# encrypt and save the updated model for next round
citadel.save(updated_model, os.environ[MODEL_KEY])
\end{lstlisting}

As shown in Pseudocode 1-3, $N$ masks $m_0, m_1, ..., m_{N-1}$ are
generated for $N$ training enclaves by the admin enclave trusted by
all secret owners (i.e., data and model owners). These masks are of the same size as the model gradients, with a sum being zero: $\sum_{i=0}^{N-1}{m_i}=0$. 
After the training starts, each training enclave first downloads and decrypts
a fresh model from the storage service, and then computes gradients with the
code shared and verified by data owners. The training enclave $i$ then
requests admin enclave for the mask $m_i$, applies it to its gradients, and
finally sends them to the aggregator enclave. The aggregator enclave collects
the masked gradients from all training enclaves, aggregates them, and updates
the model using the specified model update method. As an individual update
from each training enclave is obscured with a random mask, the model owner's
private model handling code cannot infer any information about the training data. By aggregating all
updates together, the masks cancel each other out, resulting in
the same aggregated update as it would have been without masking. 
The security of this
approach is based on the fact that if data owners' values have uniformly
random pairwise masks added to them, then the resulting values appear uniformly
random, conditioned on their sum being equal to the sum of data owner's
values, as proven in~\cite{bonawitz2017practical}.

\zhangrev{
Our centralized zero-sum masking approach protects the model confidentiality,
while guaranteeing the same level of privacy for data owners as in secure
aggregation~\cite{bonawitz2017practical}, but without the time-consuming
synchronous distributed mask generation.
Although the codes in training enclaves are shared with data owners, the secrets are omitted; thus, protected.
Static secrets like model weights, encryption keys, and training batch size are shared as placeholders, and are only dynamically loaded as environment variables once the enclaves pass attestation.
The aggregator codes remain private to the model owner because the raw training data and gradients produced from individual training enclaves are no longer accessible.
Consequently, model owners retain intellectual properties like gradient manipulation and learning rate schedules.
}

\paragraph{Offline Mask Generation.}
In our zero-sum masking approach, protecting the mask confidentiality is the
key to shielding individual updates from the model owner and the cloud provider.
Therefore, the masks have to be generated within the admin enclave and
encrypted before leaving the enclave. However, with an increasing number of training enclaves, 
the compute-intensive mask generation and 
bandwidth-intensive mask distribution
would inevitably make the admin enclave a performance bottleneck. 

To address this problem, we choose to generate masks \emph{offline} and
\emph{offload} mask distribution to the untrusted storage service after encrypting the generated masks. Before the training
starts, the admin enclave generates sufficient sets of $N$ masks, encrypts each mask with a separate key and stores
them in the storage service. During training, upon receiving a
masking request from a training enclave, the admin enclave redirects the
request to the storage service, and provides the training enclave with the
corresponding decryption key. As such, the heavy tasks of masking are removed from the critical path.

To handle potential training enclave stragglers, \citadel optionally employs a
relaxed consistency model like SSP~\cite{ho2013more}: assuming the first $K$
out of $N$ training enclaves would participate in the aggregation,
the admin enclave can return the sum of the remaining pre-generated masks
$\sum_{i=K}^{N}{m_i}$ to enclave $K$, thus ensuring the overall sum remains zero.
\subsubsection{Hierarchical Aggregation.}

The zero-sum masking solution requires \emph{one-to-all} and
\emph{all-to-one} synchronization in the mask distribution (between the
admin enclave and training enclaves) and aggregation (between training
enclaves and the aggregator enclave) phases. As more training enclaves
run in the system, such synchronization overheads become increasingly
prominent. In fact, given SGX's memory limitations, neither generating a
large number of masks within an admin enclave nor aggregating large
updates within an aggregator enclave scales.

To address this potential issue at large scale, we propose to establish 
a \emph{tree-structured hierarchical aggregation}
among training enclaves. Since our goal is to protect individual updates
from being learned by the aggregator enclave, we can utilize training enclaves
to aggregate the intermediate results, which are trusted by data owners. As
described in Algorithm~\ref{alg:citadel_recursive}, after processing a batch,
each training enclave holds its own gradients and follows a tree-structured
hierarchical aggregation scheme to accumulate gradients. 
\citadel does not require ring-reduce like communication as the update propagation pattern is all-to-one instead of all-to-all.

Assume there are $N$
training enclaves, and each \emph{leader} in the aggregation tree has $C$ children
($C-1$ neighbors need to transfer their updates to the leader in one round).
It requires $\lceil log_{C}N\rceil+1$ rounds of aggregation (height of the
aggregation tree), and on the $l$\textsuperscript{th} level, there are $\lceil
...\lceil \lceil N/C\rceil/C\rceil...\rceil$ active nodes remaining. On the
$l$\textsuperscript{th} level, we denote the $i$\textsuperscript{th} remaining
active node by ${id}^{l}_{i}$, so that each of these active nodes sends
its aggregated results from last round to a leader node ${id}^{l}_{\lfloor
N/C\rfloor}$. 
The recursion continues until the last leader accumulates the final results and sends
it to the aggregator enclave.
Hierarchical aggregation avoids the expensive all-to-one synchronization, 
eliminating communication hotspots.

\begin{algorithm}[!t]
    \caption{\citadel with Hierarchical Aggregation}
    \label{alg:citadel_recursive}
    \footnotesize
    \underline{\textbf{Training Enclave $i$:}}
    \begin{algorithmic}[1]
    \Function{StartsTraining}{}
      \For{epoch $e=0,1,2,...,E$}
          \ForAll{training batch $t=0,1,2,\cdots,T$}
            \State{Download fresh model $model$, compute gradients $g_i^{(e,t)}$}
            \State{Call \textproc{HierarchicalAggregate} to start aggregation.}
          \EndFor
      \EndFor
    \EndFunction
    \Function{HierarchicalAggregate}{}
      \If{I am leader in this round.}
        \State{Collect and accumulate intermediate results.}
        \If{I am the final leader.}
          \State{Send aggregated result $G^{(e,t)}$ to aggregator.}
        \Else
          \State{Call \textproc{HierarchicalAggregate} to send results to next leader.}
        \EndIf
      \Else
        \State{Send results to the leader.}
      \EndIf
    \EndFunction
    \end{algorithmic}
    

\end{algorithm}

\begin{figure}
    \centering
    \includegraphics[width=0.8\columnwidth]{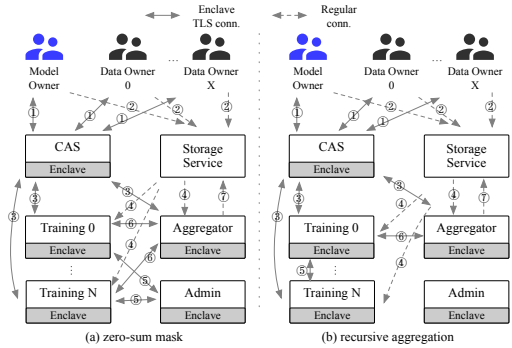}
    \vspace{-.1in}
    \caption{The workflow of \citadel with zero-sum masking. Enclave TLS connections terminate within enclaves.}
    \label{fig:flow_mask}
    \vspace{-.1in}
\end{figure}

\subsubsection{Comparison of Two Approaches.}

Both zero-sum masking and hierarchical aggregation effectively shield
the raw updates from the aggregator enclave. Zero-sum
masking requires an all-to-one communication from all training enclaves, and
then all the updates have to be aggregated in the EPC-limited aggregator
enclave, so the overall overhead grows as more training enclaves run in the
system. 

Hierarchical aggregation, on the other hand, breaks the all-to-one communication
pattern into a hierarchical aggregation tree. Although the potential network
congestion is mitigated, extra cryptographic operations are needed to protect
the communication connections on the aggregation tree. However, it is
difficult, if not impossible, to quantitatively justify such trade-off in
this scenario, as the time needed to finish a certain operation within an
enclave depends on both the memory footprint and the memory access pattern. 
Assume there are $N$ training enclaves. Let $t_{net}(x)$, $t_{enc}(x)$, and $t_{dec}(x)$ respectively
denote the time needed to transfer, encrypt, and decrypt message $x$.
Let $t_{mask}$, $t_{train}$, and $t_{apply}$ be the computation time needed to apply a mask, generate gradients, and apply gradients to model, respectively, and $t_{agg}(k)$ the time spent on aggregating updates from $k$ training enclaves.
The iteration time for zero-sum masking $t_{mask}(N)$ is estimated as
\useshortskip
\begin{equation*}
\label{eq:mask}
\small
\begin{aligned}
  t_{mask}(N) = t_{train} + t_{net}(m) + t_{dec}(m) + t_{mask} + t_{enc}(g) \\
  + t_{net}(g) + t_{dec}(g) + t_{agg}(N) + t_{apply},
\end{aligned}
\end{equation*}
where $m$ and $g$ stand for a set of mask and gradients, respectively.
Assuming each node in the aggregation tree has $C$ children, the iteration time for hierarchical aggregation $t_{tree}(N, C)$ is estimated as
\useshortskip
\begin{equation*}
\label{eq:recursive}
\small
\begin{aligned}
  t_{tree}(N, C) = (t_{enc}(g) + t_{dec}(g) + t_{agg}(C) + t_{net}(g)) \times (\lceil log_{C}N\rceil+1)  \\ + t_{train} + t_{apply}.
\end{aligned}
\end{equation*}

As a general guideline, zero-sum masking tends to work better on \emph{smaller
models} with \emph{fewer} training enclaves, as the memory footprint within
the aggregator is smaller and network congestion is less likely. When there is
a large number of training enclaves, hierarchical aggregation becomes more
favorable. We will evaluate the two approaches in \S\ref{sec:eval}.

We stress that both zero-sum masking and hierarchical aggregation require no change to distributed training algorithms, thus introducing no model accuracy loss on this front.
\citadel performs extra floating point additions during zero-sum masking and hierarchical aggregation compared to vanilla distributed training, leading to tiny numerical errors due to the exponent-alignment.
Extensive study shows that such small errors do not affect ML accuracy~\cite{smith2020generalization}.

\vspace{-2mm}
\subsection{\citadel Workflow}

Putting it all together, we elaborate on \citadel's workflow with the two aggregation 
approaches.
The workflow of zero-sum masking is depicted in Fig.~\ref{fig:flow_mask}a.
A training iteration is broken down as follows.
\Circled{1} Model and data owners attest CAS, and share their encryption keys with CAS.
\Circled{2} Model and data owners upload their encrypted model and data to storage service.
\Circled{3} CAS attests training and admin enclaves on behalf of data owners, then shares data encryption keys with training enclaves. CAS also attests training, admin, and aggregator enclaves on behalf of model owner, then shares the model decryption key to training and aggregator enclaves.
\Circled{4} Training enclaves fetch corresponding data and model, decrypt them and compute gradients; aggregator downloads and decrypts model.
\Circled{5} Training enclaves ask admin enclave for masks and have the requests redirected to
storage service with mask decryption keys.
\Circled{6} Training enclaves fetch masks from storage service, decrypt and apply them to updates, and send masked updates to aggregator.
\Circled{7} Aggregator collects masked updates, summarizes them, and updates global model. The model is then encrypted and uploaded to storage.

Similarly, Fig.~\ref{fig:flow_mask}b depicts the workflow of hierarchical aggregation, where
Steps \Circled{1}-\Circled{4} are the same as masking approach.
\Circled{5} Training enclaves hierarchically aggregate all updates until the final sum of all updates is available at training enclave 0.
\Circled{6} The aggregator enclave receives the final update and applies it to the model.
\Circled{7} The model is then encrypted and uploaded to storage service.

\paragraph{Security Against Attacks.}
\citadel's security stems from three levels.
The first level of security is provided by Intel SGX.
By verifying training enclaves and Citadel codes, data owners are ensured that data cannot be abused nor recovered.
As the second level, we reduce the possibility of data leak by splitting data handling and the model update parts.
This split ensures that model owners cannot access raw gradient updates.
It also ensures that model owners' secrets are protected from data owners, as the model update code remains private.
For the third level, placeholders (instead of actual values) are added in data handling codes shared with data owners, secrets such as model weights are protected.

\zhangrev{
SGX and its attestation ability enable \citadel
to prevent the scenario described in \cite{bonawitz2017practical} where the model owner forms a \emph{collusion} with some data owners to steal other data owners' data.
Such attacks require additional codes to be executed in the enclaves, and can be identified in the code verification process by the non-colluding data owners.
Model inversion attacks~\cite{fredrikson2015model} from data owners are avoided compared with conventional methods like FL, because the code running in training enclaves is verified by
both data and model owners and is strictly enforced by SGX.
Data owners can no longer insert backdoors into the global model by tampering updates, or perform membership inference attacks~\cite{shokri2017membership}, or conduct data extraction attacks~\cite{fredrikson2015model,hitaj2017deep},
because data owners have no access to the intermediate training results and the model, which only exist in enclaves.
}

\section{Implementation}
\label{sec:imp}

In this section, we describe the implementation details of \citadel.  We base
our implementation on SCONE~\cite{arnautov2016scone}, but it can also be
extended to other SGX-enabling frameworks such as
Graphene~\cite{tsai2017graphene} and Ryoan~\cite{hunt2018ryoan}. We use
MongoDB~\cite{mongoDB} as the storage service, and containerize all system
components and orchestrate them in Kubernetes~\cite{kubernetes}. Our implementation
consists of 5,000 lines of Python code and Linux Shell script, and is
open-sourced.\footnote{The source code of \citadel is available at~\cite{citadel-repo}.}

\paragraph{Trusted Computing Base (TCB).}
For an easy support of SGX and multiple enclave orchestration, we adopt
SCONE~\cite{arnautov2016scone} in our system. SCONE provides SGX-protected
Linux containers, so that we can utilize tools like Docker~\cite{docker} and
Kubernetes~\cite{kubernetes} to orchestrate enclaves. 
\zhangrev{
The SGX provided function \texttt{sgx\_read\_rand} is used to generate randomness, and the attacks~\cite{ragab2021crosstalk} on such generators are beyond the scope of \citadel.
}
\zhangrev{
In order to establish trust, users have to verify codes running in the enclaves.
Such verification is non-trivial, and requires some domain knowledge.
However, since \citadel and most ML toolchains are open-sourced, we believe wider adoption and public scrutiny can offload individual user's verification burden greatly.
Besides, research work like \cite{gregor2020trust} is making SGX trust management easier and more accessible.
}

\paragraph{Efficient Encryption \& Decryption.}
As the host infrastructure is not trusted, encrypted data and models must be
decrypted within the enclaves. In addition, network connections between enclaves must be secured with TLS. These requirements result in substantial cryptographic
operations performed inside an enclave. Especially during the aggregation
process, a single enclave has to decrypt results from multiple enclaves and
add them up. 
Therefore, the efficiency of cryptographic inside an enclave plays
an important role in overall performance of \citadel.

In a native setting without SGX, one way to increase performance is to
increase the parallelism with multi-processing or multi-threading. However,
inside an SGX enclave, each process runs inside its own enclave, so launching new processes is
extremely slow as it requires to set up new enclaves and initialize EPC pages.
Furthermore, the new sub-process enclaves contend with the parent enclave for
EPC, resulting in performance degradation for all of them. Our experimental
evaluations with the OpenSSL implementation of \texttt{AES-256-CBC} shows
that, encrypting and decrypting 16 AlexNet~\cite{krizhevsky2012imagenet}
models with multi-processing is at least 2X slower than processing
them serially without multi-processing in SGX. On the other hand,
SCONE~\cite{arnautov2016scone} provides efficient \emph{user-level threading}
to avoid costly system calls, so it is possible for us to improve
cryptographic operations with multi-threading. However, due to Python's Global
Interpreter Lock (GIL)~\cite{gil}, only one Python thread can run at any given
time even with multi-threading. To overcome this hurdle, we implement
our cryptographic operations in C++ and compile it with C Foreign Function Interface (CFFI)~\cite{cffi}.
This not only allows us to bypass the GIL limitation, but also exploits the
highly efficient performance of native code.

\section{Evaluation}
\label{sec:eval}

In this section, we evaluate the performance of \citadel with representative
ML models trained on a public cloud. We first examine the scalability
of \citadel with zero-sum masking in clusters of various sizes. We
then evaluate hierarchical aggregation with different configurations to quantify
how avoiding all-to-one communication helps improve system scalability.
Finally, we assess the system overhead of our design by comparing \citadel with three baselines: Chiron~\cite{hunt2018chiron}, the single-enclave approach and
native \citadel without SGX.

\subsection{Methodology}
\label{sec:eval_method}

\paragraph{Settings.}
We consider a distributed ML setting where all instances are located within
the same cluster.
\zhangrev{
We do not consider geo-distributed training, as \citadel enables data to be centralized securely in a verifiable manner, eliminating the necessity of geo-distribution like FL.}
We conduct all experiments on Azure confidential computing instances
with SGX support in \texttt{Canada Central} region with the instance \texttt{Standard\_DC4s\_v2}, which has 4 vCPUs, 16 GB of memory, and 112 MB
of EPC memory. We deploy exactly one enclave on each instance to avoid EPC contention.
To emulate multiple data owners, we randomly partition these datasets into multiple shards and encrypt them with different keys before uploading them to \citadel.
The scale of our evaluation is limited to 34 such instances (including training, aggregator and admin enclaves), because Azure limits the total number of \texttt{DCsv2} family vCPUs for non-enterprise users.
Nevertheless, we believe the trend demonstrated in our evaluation applies to
larger scale, and is sufficient to validate our implementation.

\paragraph{Benchmarking Models.}
We have implemented four ML models with their respective privacy requirements, using TensorFlow v1.15.
The first two, AlexNetS and AlexNetL, belong to the same application where a certain number of hospitals collaborate with a medical tech company to train a Retinopathy diabetes diagnosis model~\cite{retina_dataset}.
The input images are scaled to $32\times32\times3$ for AlexNetS and $96\times96\times3$ for AlexNetL, respectively.
AlexNetS has 1.25M trainable parameters while AlexNetL has 15.9M trainable parameters.
The third one SpamNet is a spam filtering model utilizing LSTM~\cite{hochreiter1997long} network with 9.6k trainable parameters, where SMS messages~\cite{sms_dataset} is input data. Here,
the model is required to be private and the SMS messages are sensitive.
The last one MNIST is a 12-layer CNN handwriting recognition model trained with MNIST dataset~\cite{lecun1998gradient}.
MNIST model has 887.5K trainable parameters.
The model owner wants to protect its intellectual property, while data owners want to remain anonymous as the adversary may forge their handwriting.

The above four workloads are backed by DL models of various sizes and cover diverse tasks.
Note that the models and input data dimensions used to evaluate SGX systems in literature are usually of modest sizes, due to memory limits and overhead of SGX's EPC.
For example, secureTF~\cite{quoc2020securetf} only adopts MNIST dataset in distributed evaluation across 3 servers;
Chiron only has simulation-based evaluation, and the biggest model investigated is AlexNet with 6 million parameters.
Compared with the state-of-the-art, our evaluations brought the training to include much larger models with realistic inputs at an industrial scale.


\paragraph{Baselines.}
We use three baselines for comparison, 
Chiron~\cite{hunt2018chiron}, the \texttt{single-enclave} approach described in \S\ref{sec:singe_enclave}, and \texttt{native-distributed} that runs
\citadel natively without SGX.
Compared with these baselines, 
we show how \citadel provides strong privacy and confidentiality, while still achieving high throughput.


\subsection{Effectiveness of Zero-Sum Masking}
\label{sec:effectiveness_of_mask}

\begin{figure}
    \centering
    \label{fig:breakdown_mask}
    \subfloat{
        \centering
        \includegraphics[width=0.7\columnwidth]{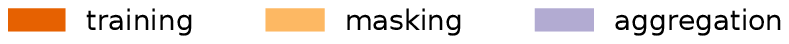}
        \label{fig:breakdown_mask_legend}
    }
    \vspace{-3.5mm}
    \setcounter{subfigure}{0}
    \subfloat[AlexNetS]{
        \includegraphics[width=0.43\columnwidth]{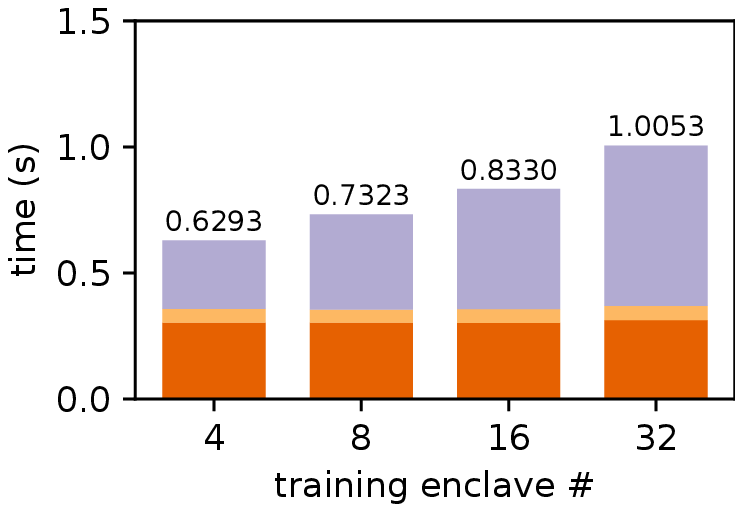}
        \label{fig:breakdown_mask_alexnet_small}
    }\hfil
    \subfloat[AlexNetL]{
        \includegraphics[width=0.43\columnwidth]{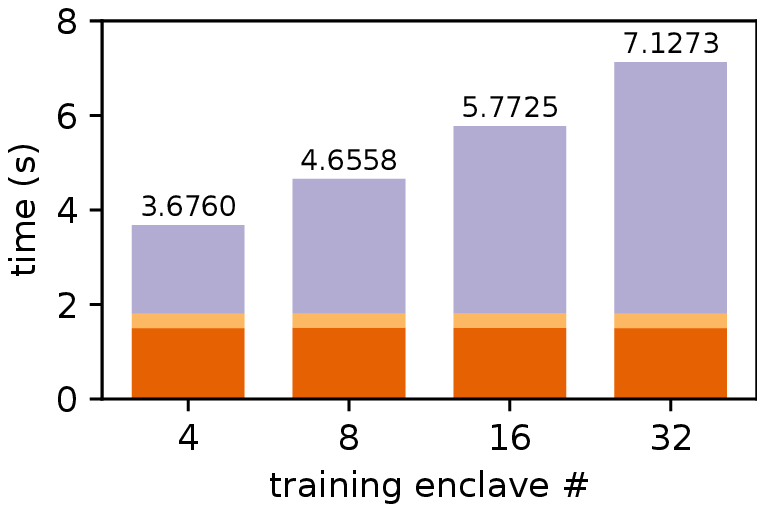}
        \label{fig:breakdown_mask_alexnet_large}
    }
    \vspace{-4mm}
    \subfloat[SpamNet]{
        \includegraphics[width=0.43\columnwidth]{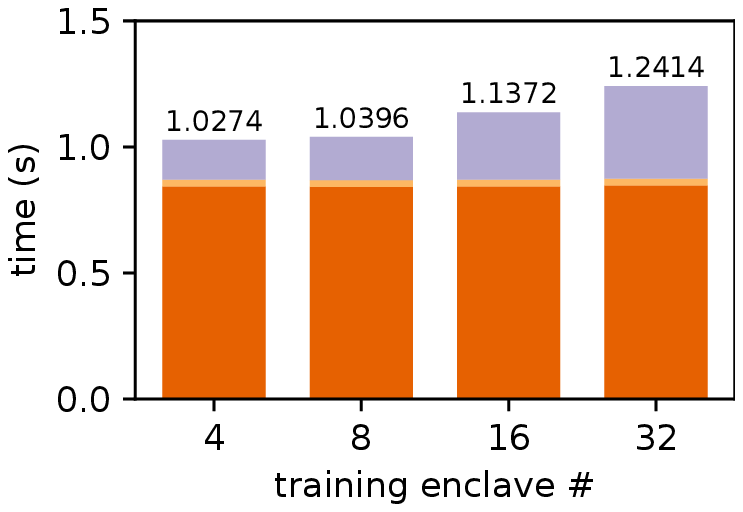}
        \label{fig:breakdown_mask_spamnet}
    }\hfil
    \subfloat[MNIST]{
        \includegraphics[width=0.43\columnwidth]{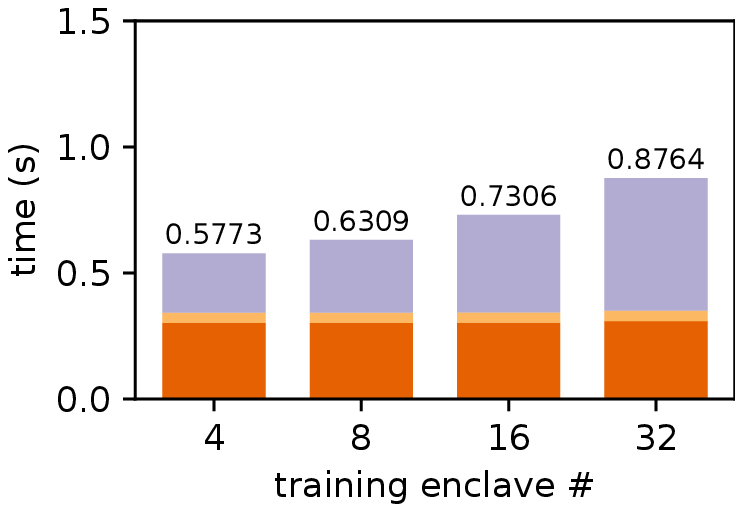}
        \label{fig:breakdown_mask_mnist}
    }
    \vspace{-.1in}
    \caption{The iteration breakdown w.r.t. training enclave numbers with zero-sum masking adopted.}
    \vspace{-.1in}
    \end{figure}

We first evaluate the effectiveness of \citadel's zero-sum masking technique.
We report the iteration time breakdown in Figs.~\ref{fig:breakdown_mask_alexnet_small}-\ref{fig:breakdown_mask_mnist}.
The iteration time is measured as the timespan from downloading fresh models in training enclaves until the aggregator enclave uploads the updated model.
Specifically, the \texttt{training} portion refers to the time spent inside training enclaves, but excludes mask-related operations.
The \texttt{masking} portion includes the time spent on requesting, downloading, decrypting and applying the masks.
The \texttt{aggregation} portion covers the time spent in the aggregator enclave, and the time used to receive all masked updates.
All results are averaged across all enclaves over multiple iterations.

As we can see, \citadel with zero-sum masking scales well with increasing number of training enclaves.
Even increasing training enclaves from 4 to 32, the overall iteration time only increases by $59.7\%$ for AlexNetS, $93.9\%$ for AlexNetL, $20.8\%$ for SpamNet, and $53.5\%$ for MNIST.
Looking into each portion, we see that: 1) training time stays constant when \citadel scales out as training operations are irrelevant to cluster size, 
2) masking time stays constant thanks to offline mask generation (\S\ref{sec:masking}), and 3) aggregation time increases (inevitably) because aggregation involves all-to-one communication and the summing-up of all masked gradient updates.
Altogether, these results show only a modest increase of iteration time as the cluster size grows, 
indicating that \citadel can accommodate a large number of data owners and complex models with reasonable performance overhead.


\subsection{Hierarchical Aggregation}
\label{sec:effectiveness_of_tree}

\begin{figure}
    \centering
    \label{fig:breakdown_tree}
    \subfloat{
        \centering
        \includegraphics[width=0.8\columnwidth]{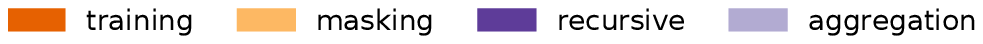}
        \label{fig:breakdown_tree_legend}
    }
    \vspace{-3mm}
    \setcounter{subfigure}{0}
    \subfloat[AlexNetS]{
        \includegraphics[width=0.43\columnwidth]{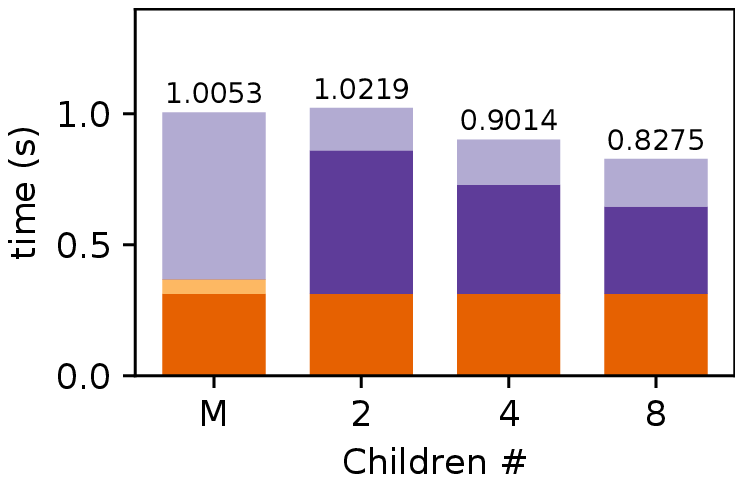}
        \label{fig:tree_alex_small}
    }\hfil
    \subfloat[AlexNetL]{
        \includegraphics[width=0.43\columnwidth]{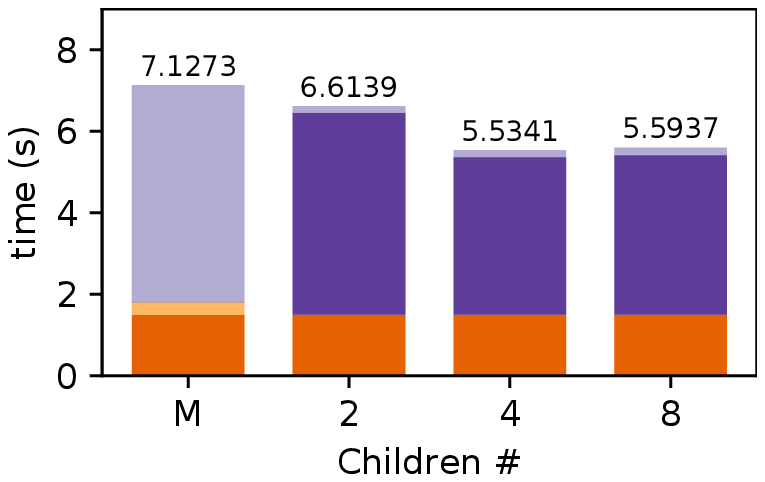}
        \label{fig:tree_alex_large}
    }
    \vspace{-3mm}
    \subfloat[SpamNet]{
        \includegraphics[width=0.43\columnwidth]{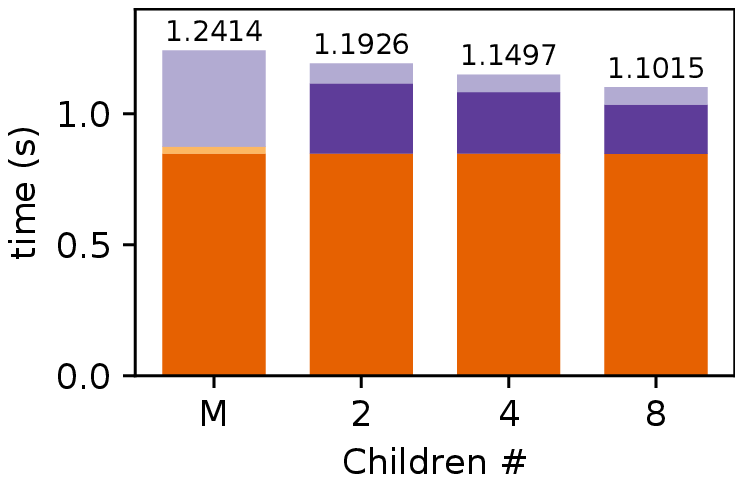}
        \label{fig:tree_spamnet}
    }\hfil
    \subfloat[MNIST]{
        \includegraphics[width=0.43\columnwidth]{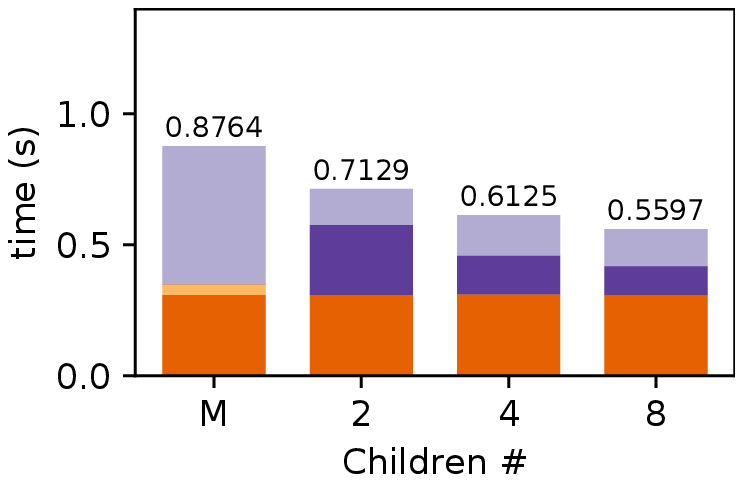}
        \label{fig:tree_mnist}
    }
    \vspace{-2mm}
    \caption{The hierarchical aggregation iteration time breakdowns w.r.t. children number. The zero-sum masking results with 32 enclaves are shown as $M$ bars.}
    \vspace{-.1in}
    \end{figure}

Although \S\ref{sec:effectiveness_of_mask} exhibits that \citadel with zero-sum masking can effectively increase throughput by adding more training enclaves, we also notice the significant aggregation overhead with increased cluster size.
In this subsection, we evaluate \citadel's hierarchical aggregation approach, and validate if it can reduce the aggregation overhead.
The results are shown in Figs.~\ref{fig:tree_alex_small}-\ref{fig:tree_mnist}.
We target the scenario with 32 training enclaves which is the largest cluster we are able to run in Azure. 
We test hierarchical aggregation with aggregation tree children set to 2, 4 and 8, and use zero-sum masking approach for reference.
The \texttt{recursive} portion in the breakdown refers to the timespan from when the first training enclave update is ready until the final model update is aggregated.

The overall iteration time is reduced by $17.7\%$, $21.5\%$, $11.3\%$, $36.1\%$ for AlexNetS, AlexNetL, SpamNet, MNIST respectively.
With AlexNetL, \citadel performs best with 4 aggregation children.
When we reduce it to 2, even the computational overhead at each aggregation level decreases, the gain is offset by increased aggregation depth. When we increase it to 8, we face large EPC overhead at each aggregation step as 8 updates have to reside in the memory simultaneously.
On the other hand, with AlexNetS, \citadel performs best with 8 children.
In conclusion, there is no one-size-fits-all optimal value across all models.
The number of children in hierarchical aggregation provides a trade-off knob for aggregation performance.
We are unable to extend our evaluation to larger scale,
but we believe hierarchical aggregation can achieve better performance when at scale, thus addressing the bottleneck in zero-sum masking.

\subsection{\citadel vs. Other SGX Systems}


\begin{figure}
    \centering
    \includegraphics[width=0.8\columnwidth]{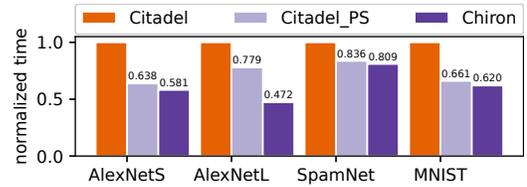}
    \caption{Chiron's runtime normalized by \citadel's. `\citadel\_PS' shows \citadel's performance with PS.}
    \vspace{-2mm}
    \label{fig:ps_comparision_ann}
    \vspace{-.35in}
\end{figure}

Chiron~\cite{hunt2018chiron} and secureTF~\cite{quoc2020securetf} are two SGX-based systems that support multi-enclave training.
They adopt parameter server (PS) as the distribution strategy, and have stronger assumptions and weaker privacy guarantees than \citadel (\S\ref{sec:multi-enclave}).
Unfortunately, neither of the systems is opensourced. As a result, we are only able to use profiling results and back-of-the-envelope calculations to showcase
how much overhead \citadel introduces in return for stronger privacy.
We use Chiron as the representative system, as secureTF shares the same distribution design as Chiron.
We use the results obtained with 32 training enclaves utilizing masking techniques, as \citadel faces the most overhead here in our evaluation.
We follow Chiron's workflow, and configure its parameter server with 8 servers and 32 workers.
The normalized iteration runtime is shown in Fig.~\ref{fig:ps_comparision_ann}, the overhead introduced by \citadel accounts for $19.1\%$ to $52.8\%$ of the runtime.
Chiron's multiple aggregator implementation reduces only half the elapsed time. However, besides offering weaker privacy guarantees, it uses 8 times the aggregation resources compared with \citadel.
Such time reduction is non-linear to the added computing power because of SGX EPC's performance constraint.
Note that \citadel can adopt PS-style multi-enclave aggregation as well. 
Such a modification is not 
essential for our privacy objectives and will require substantial engineering effort; therefore, we do not include it in \citadel yet.
Instead, we add a second bar in Fig.~\ref{fig:ps_comparision_ann} to demonstrate the \emph{expected} runtime with a PS implementation, where the remaining overhead stems from masking.


\subsection{\citadel vs. Single Enclave}
\label{sec:single_baseline}

\begin{figure}
    \centering
    \label{fig:throughput_native}
    \subfloat[AlexNetS]{
        \includegraphics[width=0.43\columnwidth]{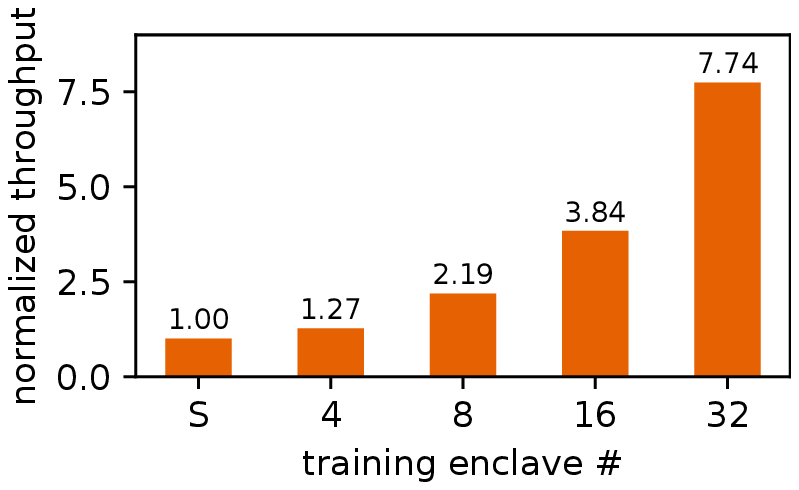}
        \label{fig:throughput_alexnet_small}
    }\hfil
    \subfloat[AlexNetL]{
        \includegraphics[width=0.43\columnwidth]{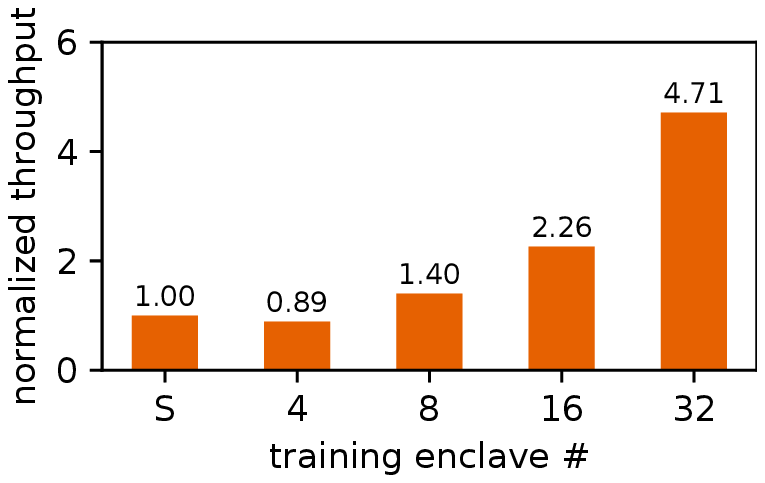}
        \label{fig:throughput_alexnet_large}
    }
    \vspace{-3mm}
    \subfloat[SpamNet]{
        \includegraphics[width=0.43\columnwidth]{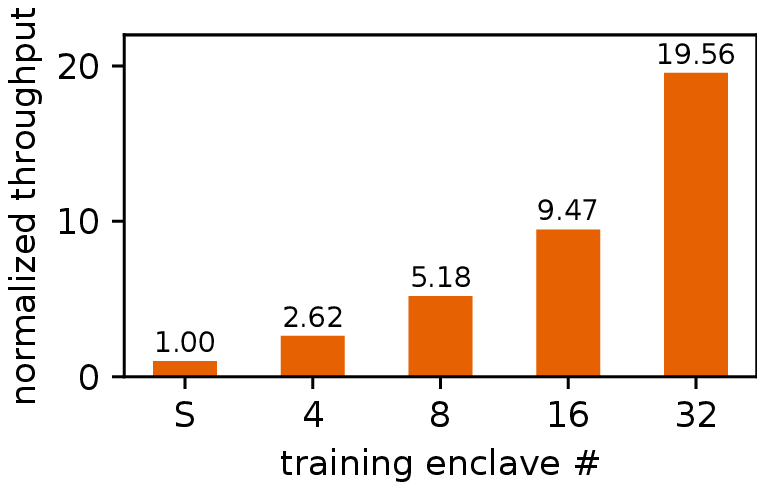}
        \label{fig:throughput_spam}
    }\hfil
    \subfloat[MNIST]{
        \includegraphics[width=0.43\columnwidth]{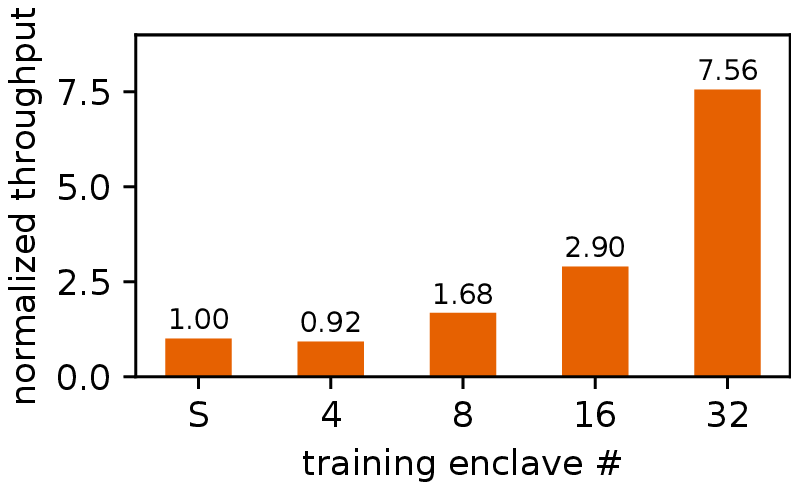}
        \label{fig:throughput_mnist}
    }
    \vspace{-.125in}
    \caption{The total throughput normalized by the \texttt{single-enclave} solution's throughput (labeled as S) w.r.t. training enclave number.}
    \vspace{-.15in}
    \end{figure}

The single-enclave solution described in \S\ref{sec:singe_enclave} can achieve data privacy and a limited protection of model confidentiality.
In this subsection, we compare \citadel with this single-enclave solution, and show that \citadel outperforms it in both privacy guarantees and performance.
Specifically, we profile the single-enclave solution's throughput via training the same models on the same Azure instance. The results are shown in Figs.~\ref{fig:throughput_alexnet_small}-\ref{fig:throughput_mnist}.
Note that we show the better results 
of zero-sum masking and hierarchical aggregation approaches.

Going from a single-enclave solution to a distributed system across multiple servers, \citadel introduces secured connections that require both network communication and cryptographic operations.
As a result, we see marginal improvements compared with \texttt{single-enclave} when using only 4 training enclaves.
However, with more training enclaves, \citadel is able to improve throughput substantially.
Note that the benefits of distributed training is more prominent for ML models with longer total training time (e.g., SpamNet in Fig.~\ref{fig:breakdown_mask_spamnet}). 
Furthermore, we aggregate model updates after each iteration in our experiments, the result therefore demonstrates the lower bound of our improvement.
One can easily improve the training performance via less communications, a.k.a., local update SGD~\cite{lin2018don,wang2018adaptive,haddadpour2019local}.
With such techniques, \citadel's throughput could be further improved.

\subsection{SGX Overhead in \citadel}
\label{sec:native_baseline}

\begin{table}
\caption{The slowdowns of \citadel. The 32-R column shows hierarchical aggregation, the rest shows zero-sum masking.}
\vspace{-2mm}
\label{tbl:slowdowns}
\centering
\footnotesize
\begin{tabular}{c|c|c|c|c|c}
\hline
\textbf{\# Training Enclaves}    & 4    & 8    & 16   & 32   & 32-R \\ \hline \hline
\textbf{AlexNetS} & 1.22 & 1.18 & 1.23 & 1.24 & 1.40 \\ \hline
\textbf{AlexNetL} & 1.09 & 1.23 & 1.44 & 1.73 & 1.65 \\ \hline
\textbf{SpamNet}  & 1.21 & 1.21 & 1.22 & 1.19 & 1.26 \\ \hline
\textbf{MNIST}    & 1.15 & 1.15 & 1.14 & 1.15 & 1.17 \\ \hline
\end{tabular}
\vspace{-.1in}
\end{table}

Finally, we compare \citadel against running at the native speed.
To do that, we repeat our evaluation on \citadel with the four workloads 
outside of SGX enclaves.
All the experiments are conducted on the same Azure Kubernetes cluster but with native docker containers running the same code as in \S\ref{sec:effectiveness_of_mask}-\ref{sec:effectiveness_of_tree}.
We seek to show how much slowdown SGX induces in the entire workflow.
We run the experiments over multiple iterations and compile \citadel's slowdown with different numbers of training enclaves in Table~\ref{tbl:slowdowns}.
The slowdown ranges from $1.09\times$ to $1.73\times$.
We show that SGX results in $15\%$--$73\%$ performance slowdown, depending on the models and scales.
With larger models like AlexNetL, memory consumption is higher, so the EPC paging happens more often, causing higher overhead.
We also notice that, the more training enclaves there are, the more memory it needs to finish aggregation, thus a higher slowdown at larger scale.


\section{Conclusion and Discussion}
\label{sec:discussion}

We presented \citadel, the first scalable system for
collaborative machine learning that protects both data privacy and model
confidentiality. \citadel partitions training into two
parts, the verifiable data handling code running in training enclaves and
the private model handling code running in the aggregator enclave. \citadel
imposes a barrier between the two parts by zero-sum masking to prevent data and model leakage, and uses hierarchical aggregation to scale up aggregation performance. \citadel is open-sourced, and the evaluation shows \citadel scales to a large number of enclaves with less than $1.7\times$ overhead despite the stringent enclave page cache (EPC) size limit.
At the end, we discuss \citadel's limitations and some future directions.



\paragraph{Large Models.} \citadel's current design does not consider models too big for single enclaves.
This issue is addressable by either increasing the EPC size with a SGX card~\cite{chakrabarti2019scaling}, or applying model parallelism to split
large models~\cite{huang2019gpipe,harlap2018pipedream,yugillis}.

\paragraph{GPU Support.} GPUs currently do not support trusted computing. Graviton~\cite{volos2018graviton} proposes an augmented GPU architecture with TEE support. In addition, Slalom~\cite{tramer2018slalom} offloads parts of the computation to GPUs with secure outsourcing.\citadel can be combined with such approaches.

\paragraph{Side Channel Attacks.} Intel SGX is currently vulnerable to side channel attacks,
\cite{ohrimenko2016oblivious} substitutes data-dependent ML operations with data-oblivious ones to address this potential issue which could be employed by \citadel.

\paragraph{Other TEEs.} \citadel is built with SGX, but its design is generally applicable to other TEE implementations including Arm TrustZone~\cite{trustzone} and Amazon Nitro~\cite{nitro}.

\paragraph{Network and Memory limits.} 
\citadel enables secure centralized ML within datacenters, thus susceptible to less critical bandwidth and memory constrains compared with FL.
Furthermore, \citadel can adopt existing techniques to reduce memory and network footprint.
It is compatible with compression schemes that support additive operations, such as sparse gradients and quantization~\cite{aji2017sparse,zhang2020batchcrypt}.


\section*{Acknowledgement}
\label{sec:ack}

We thank our shepherd Pierangela Samarati and the anonymous reviewers for
their insightful comments. We also thank Do Le Quoc (from TU Dresden and Scontain) for his generous support to set up the SCONE environment. This research 
was supported in part by ACCESS -- AI Chip Center for Emerging Smart Systems, Hong 
Kong SAR, and National Science Foundation CAREER-2048044 and IIS-1838024.

{\footnotesize
  \bibliographystyle{ACM-Reference-Format}
  \bibliography{main}
}

\end{document}